%% file: 0-main.tex
\documentclass[sigconf]{acmart}
\input{macros}

\AtBeginDocument{%
  }

 \copyrightyear{2025} 
\acmYear{2025} 
\setcopyright{acmlicensed}
\acmConference[SIGIR '25]{Proceedings of the 48th
 International ACM SIGIR Conference on Research and Development in
 Information Retrieval}{July 13--18, 2025}{Padua, Italy}
\acmBooktitle{Proceedings of the 48th International ACM SIGIR Conference on
 Research and Development in Information Retrieval (SIGIR '25), July 13--18,
 2025, Padua, Italy}
\acmDOI{10.1145/3726302.3730088}
\acmISBN{979-8-4007-1592-1/2025/07}

\begin{document}

\title{Stitching Inner Product and Euclidean Metrics for Topology-aware Maximum Inner Product Search}

\author{Tingyang Chen}
\orcid{0009-0008-5635-9326}
\affiliation{%
  \institution{Zhejiang University}
  \city{Hangzhou}
  \country{China}
}
\email{chenty@zju.edu.cn}

\author{Cong Fu}
\orcid{0000-0002-3624-6665}
\authornote{Cong Fu is the main project advisor.}
\affiliation{%
  \institution{Shopee Pte. Ltd.}
  \city{Singapore}
  \country{Singapore}}
\email{fc731097343@gmail.com}

\author{Xiangyu Ke}
\orcid{0000-0001-8082-7398}
\authornote{Corresponding author.}

\affiliation{%
  \institution{Zhejiang University}
  \city{Hangzhou}
  \country{China}
}
\email{xiangyu.ke@zju.edu.cn}

\author{Yunjun Gao}

\orcid{0000-0003-3816-8450}
\affiliation{%
  \institution{Zhejiang University}
  \city{Hangzhou}
  \country{China}
}
\email{gaoyj@zju.edu.cn}

\author{Yabo Ni}
\orcid{0000-0002-7535-8125}
\affiliation{%
  \institution{Nanyang Technological University}
  \city{Singapore}
  \country{Singapore}}
\email{yabo001@e.ntu.edu.sg}

\author{Anxiang Zeng}
\orcid{0000-0003-3869-5357}
\affiliation{%
  \institution{Nanyang Technological University}
  \city{Singapore}
  \country{Singapore}}
\email{zeng0118@ntu.edu.sg}

\renewcommand{\shortauthors}{Tingyang Chen, et al.}

\begin{abstract}
Maximum Inner Product Search (\mips) is a fundamental challenge in machine learning and information retrieval, particularly in high-dimensional data applications. 
Existing approaches to \mips either rely solely on Inner Product (\ip) similarity, which faces issues with local optima and redundant computations, or reduce the \mips problem to the Nearest Neighbor Search under the Euclidean metric via space projection, leading to topology destruction and information loss. 
Despite the divergence of the two paradigms, we argue that there is no inherent binary opposition between \ip and Euclidean metrics. By stitching \ip and Euclidean in the indexing and search algorithms design, we can significantly enhance \mips performance. 
Specifically, this paper explores the theoretical and empirical connections between these two metrics from the \mips perspective. Our investigation, grounded in graph-based search, reveals that different indexing and search strategies offer distinct advantages for \mips, depending on the underlying data topology. Building on these insights, we introduce a novel graph-based index called Metric-Amphibious Graph (\magg) and a corresponding search algorithm, Adaptive Navigation with Metric Switch ({\sf ANMS}). 
To facilitate parameter tuning for optimal performance, we identify three statistical indicators that capture essential data topology properties and correlate strongly with parameter tuning. 
Extensive experiments on 12 real-world datasets demonstrate that \magg outperforms existing state-of-the-art methods, achieving up to 4$\times$ search speedup while maintaining adaptability and scalability.
\end{abstract}




\maketitle

\input{1-intro}

\input{2-prelim}
\input{3-anaylsis}
\input{4-data}
\input{6-exp}

\input{7-related}
\input{8-conclu}

\begin{acks}
This work was supported in part by the NSFC under Grants No. (62025206 and U23A20296), Zhejiang Province’s “Lingyan” R\&D Project under Grant No. 2024C01259, Ningbo Yongjiang Talent Introduction Programme (2022A-237-G).
\end{acks}


\bibliographystyle{ACM-Reference-Format}
\balance
\bibliography{ref}


\end{document}

%% file: macros.tex
\usepackage{hyperref} 
\usepackage{graphicx}
\usepackage{amsmath}
\usepackage{amsthm}
\usepackage{amsfonts}
\usepackage{subfigure}
\usepackage{balance}
\usepackage[ruled,vlined]{algorithm2e}
\usepackage{multirow}

\newtheorem{theor}{Theorem}
\newtheorem{propsion}{Proposition}

\newtheorem{defn}{Definition}

\newtheorem{fact}{Fact}

\newcommand{\squishlist}{
 \begin{list}{$\bullet$}
  {  \setlength{\itemsep}{0pt}
     \setlength{\parsep}{3pt}
     \setlength{\topsep}{3pt}
     \setlength{\partopsep}{0pt}
     \setlength{\leftmargin}{2em}
     \setlength{\labelwidth}{1.5em}
     \setlength{\labelsep}{0.5em}
} }
\newcommand{\squishlisttight}{
 \begin{list}{$\bullet$}
  { \setlength{\itemsep}{0pt}
    \setlength{\parsep}{0pt}
    \setlength{\topsep}{0pt}
    \setlength{\partopsep}{0pt}
    \setlength{\leftmargin}{2em}
    \setlength{\labelwidth}{1.5em}
    \setlength{\labelsep}{0.5em}
} }

\newcommand{\squishdesc}{
 \begin{list}{}
  {  \setlength{\itemsep}{0pt}
     \setlength{\parsep}{3pt}
     \setlength{\topsep}{3pt}
     \setlength{\partopsep}{0pt}
     \setlength{\leftmargin}{1em}
     \setlength{\labelwidth}{1.5em}
     \setlength{\labelsep}{0.5em}
} }

\newcommand{\squishend}{
  \end{list}
}

\newcommand{\eat}[1]{}

\newcommand{\kw}[1]{{\ensuremath {\mathsf{#1}}}\xspace}

\newcommand{\stitle}[1]{\vspace{1.5ex}\noindent{\bf #1}}

\newcommand{\eetitle}[1]{\vspace{0.8ex}\noindent{\em\underline{#1}}}

\newcounter{ccc}

\DeclareMathOperator*{\argmin}{arg\,min}
\DeclareMathOperator*{\argmax}{arg\,max}

\newcommand\redout{\bgroup\markoverwith
{\textcolor{red}{\rule[.5ex]{2pt}{2pt}}}\ULon}

\newcommand{\mips}{\kw{MIPS}}
\newcommand{\gnns}{\kw{GNNS}}
\newcommand{\nns}{\kw{NNS}}
\newcommand{\lsh}{\kw{LSH}}

\newcommand{\ip}{\kw{IP}}
\newcommand{\magg}{\kw{MAG}}



%% file: 1-intro.tex
\section{Introduction}
\label{sec:intro}

Maximum Inner Product Search (\mips) is a fundamental task in machine learning and information retrieval~\cite{lewis2020retrieval, seo2019real}, particularly with the widespread use of high-dimensional vector representations based on inner product or cosine similarity. 
Key applications, including recommendation systems \cite{xu2018deep}, query-answering chatbots \cite{ahmad2019reqa}, multi-modal retrieval \cite{wang2024must}, and Retrieval Augmented Generation (RAG) \cite{asai2023retrieval}, rely on efficiently searching large vector databases to identify items that maximize similarity to a query vector.  
Fast and accurate \mips enhances system performance and user experience by leaving more time for complex model inference.

Two primary paradigms for fast \mips have emerged.
The first directly operates in the inner product space~\cite{morozov2018non, guo2020accelerating, zhao2023fargo, guo2016quantization, liu2020understanding}, constructing specialized indices. 
However, the absence of triangle inequality {\em undermines the geometric theoretical support} of efficient indexing, resulting in drawbacks like high memory cost, excessive computations, and susceptibility to local optima~\cite{tan2021norm}. 
The second paradigm~\cite{zhao2023fargo, zhou2019mobius, shrivastava2014asymmetric} reduces \mips to the nearest neighbor search (\nns) problem in a transformed Euclidean space, enabling the use of advanced \nns indices. 
However, this reduction often {\em relies on nonlinear projections and strong theoretical assumptions}~\cite{zhou2019mobius, zhao2023fargo}, which can distort data topology and cause information loss~\cite{morozov2018non}, leading to reduced performance and scalability.

These paradigms differ fundamentally in their reliance on distinct distance metrics. However, we argue that {\em Inner Product and Euclidean metrics are not mutually exclusive}. 
Focusing on graph-based retrieval~\cite{wang2021comprehensive}, our theoretical and empirical analysis (\S\ref{sec:analysis}) reveals that both Euclidean- and \ip-based strategies can enhance \mips efficiency. 
Specifically, we observe that: 
\textbf{(1)} Euclidean-based indexing and search ensure \textbf{strong connectivity and global reachability}, as established in prior studies~\cite{fu2019fast}, but typically require \textbf{longer traversal paths} to reach relevant candidates in \mips.
\textbf{(2)} \ip-based indexing and search {\bf concentrate edges toward high-norm points}, accelerating retrieval among high-\ip candidates~\cite{liu2020understanding} but suffer from reduced connectivity and local optima traps due to in-degree concentration. 
In addition, the absence of an effective edge sparsification strategy with theoretical guarantees results in high memory costs and excessive computations.

\begin{figure}[tb!]
\vspace{1ex}
\centering
\centerline{\includegraphics[width=0.92\linewidth]{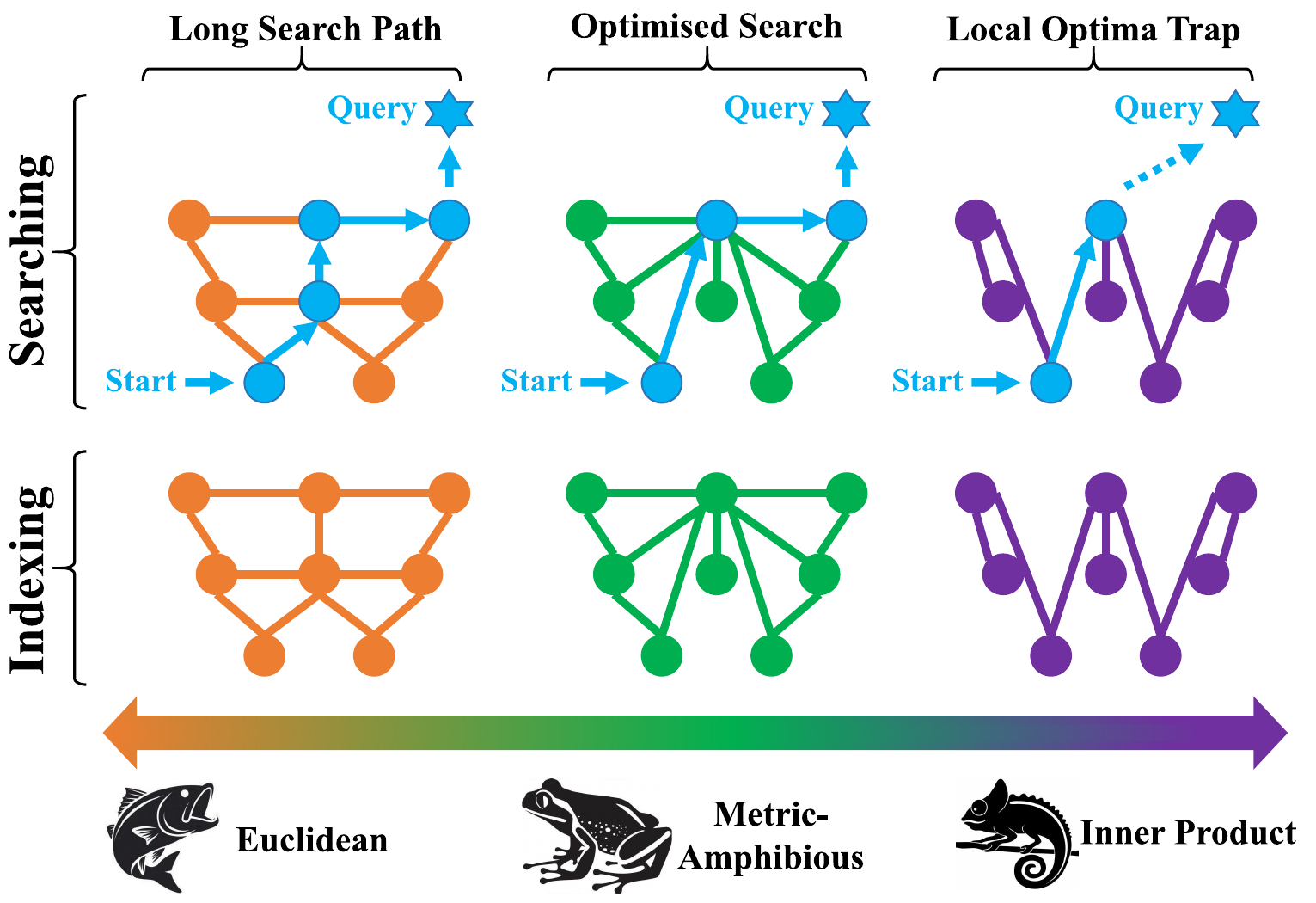}}
\vspace{-2ex}
\caption{Illustration of our motivation. Euclidean-based indexing and searching ensure strong connectivity but suffer from inefficient traversal. \ip-based indexing and search accelerates navigation toward high-relevance regions but risks getting trapped in local optima. Integrating both strategies balances connectivity, search speed, and adaptability to diverse data topologies, leading to improved performance.}
\vspace{-2ex}
\label{fig:figure1}
\end{figure}

To leverage the strengths of both paradigms while mitigating their limitations, we propose a hybrid framework that integrates Euclidean- and \ip-based strategies for indexing and search.
In the indexing phase, we introduce a novel \ip-based edge selection strategy, which is combined with the Euclidean-based approach to balance global connectivity with rapid convergence to target items. 
In the search phase, we dynamically adjust the navigation between Euclidean- and \ip-based traversal to optimize robustness and avoid local optima traps. Figure~\ref{fig:figure1} illustrates our design and motivation.

\vspace{1mm}
\noindent\textbf{Contributions.} Our contributions are highlighted as follows:

\vspace{0.3mm}
\noindent \textbf{(1) Theoretical Foundations}: We establish comprehensive connections between \ip and Euclidean metrics in the \mips setting and introduce \ip-oriented edge selection strategies with theoretical guarantees for improved efficiency.

\vspace{0.3mm}
\noindent \textbf{(2) Novel MIPS Framework}: We propose Metric-Amphibious Graph (\magg), a new framework that integrates Euclidean- and \ip-based edges, along with a novel search algorithm, namely Adaptive Navigation with Metric Switch (\textsf{ANMS}). 
Notably, our approach \textbf{DOES NOT} require space transformation, avoiding related drawbacks that plague prior methods.

\vspace{0.3mm}
\noindent \textbf{(3) Data-Driven Parameter Tuning}: We introduce three statistical indicators that capture key data topology properties, providing insights for efficient parameter tuning. These indicators enable the seamless adaption of \magg to various data distributions.

\vspace{0.3mm}
\noindent \textbf{(4) Comprehensive Experiments}: We evaluate our approach on 12 real-world datasets with varying topology, cardinality, dimensionality, and modalities. Our approach achieves up to 4$\times$ speedup compared to advanced graph-based retrieval methods.

%% file: 2-prelim.tex
\section{Preliminaries}
\label{sec:prelim}

\stitle{Notations.} Let $\mathbb{R}^d$ denote $d$-dimensional real coordinate space. $\{\cdot\}$ denote sets. $\mathcal{D} = \{ x_1, \dots, x_n \} \subset \mathbb{R}^d$ denote a vector dataset. $\langle x, y \rangle$ denotes the inner product (\ip) between vector $x$ and $y$. 
$\lVert x \rVert$ gives the Euclidean norm of vector $x$. $V_x$ denotes the Voronoi Cell under \ip metric associated with $x$. $G = (V, E)$ denotes a graph, where $V$ is the node set and $E$ is the edge set. $\sup (S)$ denotes the supremum of a set $S$. $O(\cdot)$ is the big $O$ notation.

\stitle{Problem Definition.} The \textbf{Maximum Inner Product Search (\mips)} problem is defined as: Given a query vector $q\in\mathbb{R}^d$ and a dataset $\mathcal{D}$, find the vector $x^*$ such that $x^* = \arg\max_{x \in \mathcal{D}} \langle q, x \rangle.$ Similarly, the \textbf{Nearest Neighbor Search (\nns)} problem in Euclidean space is defined as finding the vector $x^*$ such that $ x^* = \arg\min_{x \in \mathcal{D}} \lVert q - x \rVert$. Recently, researchers have focused on the \textbf{approximate \mips/\nns} problem, which allows for an acceptable loss in accuracy for faster query processing. The approximate \mips problem is defined as follows: Given a query $q \in \mathbb{R}^d$, a dataset $\mathcal{D} \subset \mathbb{R}^d$, and an approximation ratio $\epsilon \in (0,1)$, let $x^* \in \mathcal{D}$ be the exact \mips\ solution for $q$. The goal is to find a vector $x \in \mathcal{D}$ satisfying: $\langle x, q \rangle \ge \epsilon \cdot \langle x^*, q \rangle$. 

\stitle{Graph-based Indexing.} Graph-based indices have gained prominence in \nns due to their efficiency in Euclidean spaces~\cite{malkov2018efficient,fu2019fast,wang2021comprehensive}. Similarly, \mips-oriented graph-based methods often construct indices based on the \ip-Delaunay Graph, analogous to the Delaunay Graph in Euclidean space~\cite{morozov2018non}. \ip-Delaunay Graph is defined as:

\begin{defn}[\textbf{IP-Delaunay Graph}]
Given a dataset $\mathcal{D} \subset \mathbb{R}^d$, the \textbf{IP-Voronoi Cell} associated with a vector $x\in\mathcal{D}$ is $V_x=\{y\in\mathbb{R}^d\mid\langle y,x\rangle >\langle y,z\rangle , \forall z\in\mathcal{D}, z\neq x\}$. the \ip-Delaunay Graph $G$ is constructed by connecting any two nodes $x_i$ and $x_j$ with a bi-directional edge if their Voronoi cells $V_{x_i}$ and $V_{x_j}$ are adjacent in $\mathbb{R}^d$.
\label{def:ip-delaunay}
\vspace{-2mm}
\end{defn}

The \ip-Delaunay Graph is inherently densely connected, especially in high dimensions, which is inefficient for \mips. However, unlike in Euclidean space, there lacks an efficient way to sparsify the \ip-Delaunay graph to improve search efficiency and lower memory requirements, while maintaining theoretical guarantees \cite{morozov2018non}. 

\stitle{Graph-based Search.} Graph-based search follows a common iterative routine across both \ip and Euclidean metrics: (1) Start from an initial candidate set; (2) Expand search iteratively by checking neighboring nodes; (3) Update candidate pool based on proximity to the query; (4) Continue until convergence criteria are met (see Algorithm \ref{alg:gnns}). The key difference lies in the search objective. Euclidean Search minimizes $\lVert x-q\rVert$, while \mips maximizes $\langle x,q\rangle$.

\begin{algorithm}[t]
  \caption{\textsc{Greedy Search For Graphs}}
  \label{alg:gnns}
  \small
  \KwData{Graph $G$, query $q$, candidate set size $l_s$, result set size $k$, similarity measure $M(\cdot,\cdot)$.}
  \KwResult{Top-$k$ result set $R$.}
  Initialize candidate set $Q$ with size $l_s$ randomly;\\
  Configure $Q$ as heap prioritizing points closest to $q$ w.r.t $M$; \\
  \While{There exists unvisited points in $Q$}{
    $p\gets$ first unvisited point in $Q$; Mark $p$ as visited;\\
    $N_p \gets$ neighbors of $p$ in $G$; \\
    \For{each node $n$ in $N_p$}{
        $Q.insert((n,M(n,q)))$;
    }
    $Q.resize(l_s)$;\\
  }

  \textbf{return} $R \gets$ Top-$k$ points in $Q$;
\end{algorithm}

\begin{figure*}[ht]
\begin{center}
\includegraphics[width=0.93\linewidth]{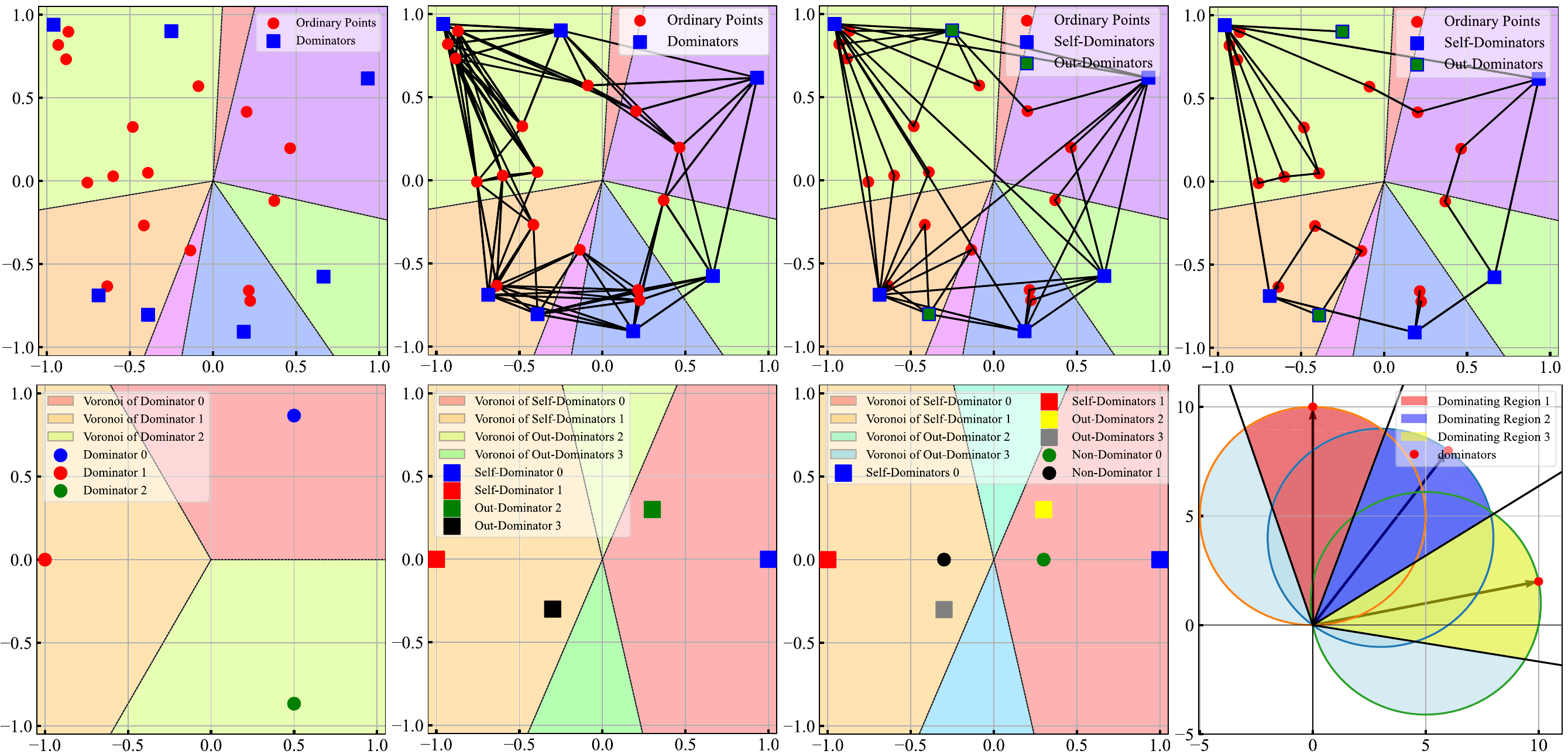}
\end{center}
\caption{Illustrations of \ip geometry concepts, simulated on small toy 2D data. (a) \ip-Voronoi cells (open hyper-cones) with associated dominators. (b) K-Maximum Inner product (K-\textsf{MIP}) Graph . (c) K-Naive Dominator Graph (NDG). (d) Optimal MAG. (e) Simpler illustration of self-dominators. (f) Showcase of out-dominators dominating vacant regions. (g) Showcase of ordinary points residing in self-dominators' Voronoi$_{ip}$ cells. (h) Valid dominating region of dominators—capped hyper-cones.} 
\vspace{-1mm}
\label{fig:main-concepts}
\end{figure*}

%% file: 3-anaylsis.tex
\section{Theoretical \& Empirical Analysis}

\label{sec:analysis}
Graph-based search efficiency is often analyzed using the formula $C = D \times L$, where $C$ represents the total computation, $D$ is the average out-degree, and $L$ is the search-path length~\cite{wang2021comprehensive}. 
Optimizing efficiency requires minimizing both $D$ and $L$, with an underlying assumption on strong graph connectivity. 
In the Euclidean space, many studies~\cite{fu2019fast,malkov2018efficient,peng2023efficient} have explored edge-sparsification strategies for Delaunay Graphs, providing theoretical guarantees on reachability and short search-path lengths. 
While in \mips context, three fundamental questions arise: \textbf{(1)} How to design a sparse graph structure under \ip metric?  \textbf{(2)} How to ensure the reachability of \mips solutions? \textbf{(3)} What are the expected search path lengths?

To address these issues and motivate this work, we propose and analyze a geometry-based domination property under the \ip metric, explore the connections between Euclidean-based \nns and \mips, and provide empirical observations to validate our theoretical analysis. These insights inform the design of our proposed solutions, which are detailed in \S\ref{sec:method}.

\subsection{Geometry Domination under Inner Product}
\label{sec:domination}
We begin by introducing dominators, tailored to the \mips problem as a key to effective graph sparsification.

\begin{defn}[\textbf{Dominators}]
   A vector $x \in \mathcal{D}$ is a dominator of its Voronoi$_{ip}$ cell $V_x$ if, for all  $y \in V_x$ and all $z \in \mathcal{D}$ with $z \neq x$, it holds that $\langle y,x \rangle > \langle y,z \rangle$. Let $S_{dom}$ represent dominators in $\mathcal{D}$.
\label{def:dominator}
\end{defn}

Intuitively, if a query vector $q$ lies in Voronoi cell $V_x$ dominated by point $x$, then $x$ is the exact \mips answer for $q$. This observation motivates the construction of the Naive Dominator Graph below:

\begin{defn}[\textbf{Naive Dominator Graph (NDG)}]
\label{def:ndg}
Given a dataset $\mathcal{D} \subset \mathbb{R}^d$, an \textbf{NDG} $G$ is constructed as follows: For each point $x_i \in \mathcal{D}$, sort the remaining points in descending order of $\langle x_i, y_j \rangle$ to form a list $L(x_i)$. Starting from the beginning of $L(x_i)$, evaluate each point $y_j$ using these conditions: \textbf{(1):} $\langle y_j, y_j \rangle \geq \langle y_j, y_k \rangle$ for all $k < j$. \textbf{(2)} $\langle y_k, y_k \rangle \geq \langle y_j, y_k \rangle$ for all $1 < k < j$. 
If $y_j$ satisfies these conditions, add a bi-directional edge $(x_i, y_j)$ to the graph $G$.
\end{defn}

Under this construction, two types of dominators emerge (illustrated in Figure~\ref{fig:main-concepts}): \textbf{(1)} $x$ is a \textbf{self-dominator} if $\forall y \in \mathcal{D}$, $\langle x,x \rangle > \langle x, y \rangle$, meaning $x$ dominates itself and resides within $V_x$; \textbf{(2)} $x$ is an \textbf{out-dominator} if $x$ is a dominator but $\exists y \in \mathcal{D}$ such that $\langle x,x \rangle \leq \langle x, y \rangle$, then $x$ is dominated by $y$ and belongs to $V_y$. We can prove the following properties for \textsf{NDG}:

\begin{theor}
\label{theorem:ndg-property}
Given a dataset $\mathcal{D} \subset \mathbb{R}^d$, (1) an NDG is a strongly connected graph; (2) $\forall x_i\in\mathcal{D}$, $x_i$ is connected to at most one out-dominator and all self-dominators in an NDG.
\end{theor}
\begin{proof}
Consider a point $x_i \in \mathcal{D}$ and the sorted list $L(x_i) = [y_1, y_2, \dots, y_m]$ that satisfies Definition \ref{def:ndg}, we then have $\langle y_1, x_i \rangle > \langle y_j, x_i \rangle, \forall j > 1$. Therefore, $y_1$ dominates $x_i$ but is not ensured to be a self-dominator, i.e., a potential out-dominator.

For any remaining point $y_j, \forall j>1$, Definition \ref{def:ndg} ensures that:

\noindent \textbf{(1)} For any pair of $y_j$ and $y_k$ with $k < j$, we have $\langle y_j, y_j \rangle \geq \langle y_j, y_k \rangle$. 

\noindent \textbf{(2)}  For any pair of $y_j$ and $y_l$ with $l > j$, we have $\langle y_j, y_j \rangle \geq \langle y_j, y_l \rangle$. 

This confirms that $\forall j > 1, y_j$ is a self-dominator, while all the other filtered points cannot be self-dominators. Consequently, each $x_i$ is linked to at most one out-dominator and all self-dominators, proving Property (2). Since all edges are bi-directional, and each node is connected to at least one self-dominator, the graph is strongly connected, proving Property (1).
\end{proof}

By Property (2) in theorem \ref{theorem:ndg-property}, \textsf{NDG}'s sparsity is determined by the number of self-dominators, which can be estimated analytically:

\begin{propsion}
\label{props:density}
Given a dataset $\mathcal{D} \subset \mathbb{R}^d$ where vectors are element-wise i.i.d. and drawn from the standard Gaussian distribution $\mathcal{N}(0,1)$, the probability that a vector $x \in \mathcal{D}$ with norm $\lVert x\rVert = r$ is a self-dominator is given by $\mathcal{P}_{dom}(x) = \Phi(r)$, where $\Phi(\cdot)$ is the cumulative distribution function (CDF) of the standard Gaussian.
\end{propsion}
\begin{proof}
By definition, $x$ of norm $r$ is a self-dominator with the probability:
$\mathcal{P}_{dom}(x) = P(\langle x, x \rangle > \langle x, y \rangle \mid \lVert x\rVert = r) \label{eq:p-dom-def}$.

Given $\lVert x\rVert = r$, $\langle x, x \rangle = r^2$ is deterministic. Since $y$ is independent of $x$ and its elements are i.i.d., conditioned on $\lVert x\rVert = r$, the inner product $\langle x, y \rangle$ becomes a linear combination of $d$ independent standard Gaussian and follows: 
$\langle x, y \rangle \mid \lVert x\rVert = r \sim \mathcal{N}(0, r^2) $

By plugging in and rearranging, equation (\ref{eq:p-dom-def}) becomes:

\hspace*{4em} $\mathcal{P}_{dom}(x) = P\left( \frac{\langle x, y \rangle}{r} < r \mid \lVert x\rVert = r \right) $

Hence by standardizing, $Z=\frac{\langle x, y \rangle}{r} \sim \mathcal{N}(0,1)$, we have:

\hspace*{4em} $\mathcal{P}_{dom}(x) = P(Z<r) =\Phi(r) $

where $\Phi(\cdot)$ is the CDF of the standard Gaussian distribution. 
\end{proof}

Proposition~\ref{props:density} confirms that \textbf{self-dominators are predominantly high-norm vectors}, consistent with prior empirical findings \cite{liu2020understanding}. 
Specifically, if a vector $x$ satisfies $||x||>4$, it is almost sure to be a self-dominator (note that "4" is derived from the $\mathcal{N}(0,1)$ assumption and may shift depending on the actual distribution’s mean and variance). 
Given that $||x||$ follows a Chi distribution under the assumptions in Proposition~\ref{props:density}, the expected number of self-dominators in a dataset $\mathcal{D}$ can be estimated as: $n\times P(\lVert x \rVert>r) = n\left(1-\frac{\gamma(d/2,r^2/2)}{\Gamma(d/2)}\right)$, where $\gamma(\cdot)$ is the lower incomplete gamma function and $\Gamma(\cdot)$ is the gamma function. 
This expression shows that {\em dominator density is also related to the dimensionality $d$}. 
Notably, $d$ should be considered as the intrinsic dimensionality of the data, which captures the effective degrees of freedom in the data and can be significantly lower than the actual dimensionality.
This, in turn, affects the expected norm distribution and the density of self-dominators.
For example, in the 784-dimensional \textsf{MNIST1M} dataset (Table \ref{tab:prop}), only 6.2\% of vectors are self-dominators, highlighting high \textsf{NDG} sparsity in certain structured data (refer to \S\ref{sec:method}). 

Despite the general sparsity of self-dominators, the \textsf{NDG} can exhibit dense connectivity in certain data distributions.
Analogous to K-\textsf{MIP} graphs, which approximate \ip-Delaunay graphs as a practical alternative (avoiding high-degree and enhancing memory efficiency~\cite{morozov2018non}), we propose \textbf{K-NDG} as an efficient approximation of \textsf{NDG}: K-\textsf{NDG} links each node only to dominators that maximize the inner product. 
Unlike K-\textsf{MIP}, which connects each node to any high-\ip neighbors, K-\textsf{NDG} restricts edges to dominators, leading to a sparser and more memory-efficient structure (see Figure~\ref{fig:main-concepts}).

\stitle{Remark 3.1. Above theoretical advancements pioneers in the literature of MIPS} and can be summarized as: \textbf{(1)} Dominators are optimal \mips answers; \textbf{(2)} All dominators are reachable in \textsf{NDG}; \textbf{(3)} K-\textsf{NDG} is a sparser and practical alternative for \ip-Delaunay.

Despite these merits, two structural challenges remain: \textbf{(1)} The norm distribution bias and sparsity of self-dominators lead to a high concentration of out-edges toward them, increasing the risk of local optima traps and inherent low-connectivity of K-\textsf{NDG}. 
\textbf{(2)} For top-K retrieval tasks, where not all solutions are dominators, K-\textsf{NDG} may struggle with reduced generalizability.

Our preliminary tests uncover these concerns. Specifically, we conduct top-K \mips experiments on two datasets, varying K in $[1,100]$. Table~\ref{tab:ndg} presents the results. 
\textbf{Key observations} are: (1) As $K$ increases, K-\textsf{NDG}’s relative speedup over the best competitor decreases significantly. 
(2) On Shopee1M, K-\textsf{NDG} underperforms the baseline as $K$ grows. At certain $K$, it even hits an accuracy bottleneck. This confirms our analysis that K-\textsf{NDG} may suffer from connectivity issues, potentially trapping search in local optima.

To tackle these issues, the next section investigates connections between \nns and \mips \textbf{without requiring space transformations} to utilize the strengths of Euclidean-based methods for \mips.

\subsection{Connect Euclidean NNS To MIPS on Graphs}
\subsubsection{Euclidean-Based Graphs Strengthens Connectivity For \mips}
\label{sec:3-2-1}

\noindent Existing nearest neighbor graphs (\textsf{NNG}s) such as \textsf{HNSW}~\cite{malkov2018efficient} and \textsf{NSG}~\cite{fu2019fast} ensure strong connectivity, retain high sparsity, and have been extensively tested at scale. 
Such advantages can benefit \mips on graphs with proper utilization. Prior work~\cite{chen2025maximum} establishes a fundamental duality between \mips and \nns, formalized as follows:

\begin{table}[tb!]
  \vspace{-1mm}
  \caption{The performance improvement of K-\textsf{NDG} relative to the best competitor for various $K$ values at 98\% recall.}
  \vspace{-3mm}
  \small
  \label{tab:ndg}
  \resizebox{\linewidth}{!}{
  \begin{tabular}{cccccc}
    \toprule
    Datasets & k=1 & k=20 & k=50 & k=100\\
    \midrule
    \textsf{Music100} & 38\%  & 30\% & 27\% & 21\% \\
    \textsf{Shopee1M} & 100\% &  -20\% & Precision  limit (0.68) & Precision limit (0.49) \\
    \bottomrule
  \end{tabular}}
  \vspace{-4mm}
\end{table}

\begin{fact}
\label{fact:scale_nn}
Given vector database $\mathcal{D} \subset \mathbb{R}^d$ and Euclidean proximity graph $G=(V,E)$, for any $q \in \mathbb{R}^d$, there exists a scalar $\bar{\mu}$ such that for all $\mu > \max(\bar{\mu},0)$, the nearest neighbor of $q'=\mu q$ in $\mathcal{D}$ aligns with the \mips solution for $q$. Furthermore, when using the standard Graph Nearest Neighbor Search (\gnns) on $G$~\cite{prokhorenkova2020graph}, the search behavior for $q$ under the \ip metric matches that for $q'$ under the Euclidean metric.
\end{fact}

This result implies that greedy search (Algorithm~\ref{alg:gnns}) under \ip metric can be performed directly on \textsf{NNG}s, making them potentially effective indices for \mips. From our perspective, this suggests that \textbf{edges selected under the Euclidean metric can complement \ip-oriented graphs, enhancing connectivity, particularly for top-K retrieval}, where solutions may not be dominators.

\subsubsection{Euclidean Oriented Navigation Avoids Local Optima in \mips}
\label{sec:3-2-2}

\noindent The \mips objective can be stated as: $\max \langle p,q\rangle=\max \lVert p\rVert \lVert q\rVert \cos\theta$. Thus, \mips consists of two key processes: norm expansion of $\lVert p\rVert$ and minimizing angle $\theta$. We now show that executing a Euclidean-oriented search on a query $q$ effectively reduces angular distance.

\begin{propsion}
Consider a vector database $\mathcal{D} \subset \mathbb{R}^d$ containing $n$ points, where $n$ is sufficiently large to ensure robust statistical properties. Assume each element of the base vectors are sampled from $\mathcal{N}(0,1)$, independently and identically distributed (i.i.d.). The angle $\theta$ between point $x$ and its nearest neighbor $y$ in $\mathcal{D}$ can be estimated by $\arccos \min\left(\left(\frac{1}{td}\left(\log n + \frac{d}{2} \log (1-t^2)^{-1}\right)\right),1\right)$, where $0 < t < 1$ is a tuning parameter. As $n \to \infty$, $\theta$ converge to $0^\circ$ .
\end{propsion}
\begin{proof}
Because $y$ is the nearest neighbor of $x$, we have $y = \argmin_z||x-z||$. Given $||x-z||=\sqrt{||x||^2+||z||^2-2\langle x,z\rangle}$, we can get $y = \argmax_z \langle x,z\rangle$ subject to $2\langle x,z\rangle < ||x||^2+||z||^2$. 

Given $x$ are i.i.d. sampled from $\mathcal{N}(0,1)$, $||x||^2$ follows a chi-square distribution with $d$ degree of freedom. According to Central Limit Theory and concentration of measure, $||x||^2$ distribution will concentrate sharply around $d$, making $d$ a good estimation of $\lVert x\rVert$. By Substituting, we can get $y \approx \argmax_y \langle x,y\rangle, s.t. \langle x,y\rangle < d$.

Let $M=\max \langle x,y\rangle$, which follows a distribution characterized by a modified Bessel function of the second kind~\cite{bowman1958introduction}, given $x,y$ are i.i.d sampled from $\mathcal{N}(0,1)$. Using Extreme Value Theory, we can estimate $\max \langle x,y\rangle$ below. With Jensen's Inequality~\cite{mcshane1937jensen}, we have:
\begin{equation}
    \nonumber
    \begin{aligned}
        e^{t\mathbb{E}[M]} &\le E[e^{tM}] \le nMGF(t)_{M}  \\
        \mathbb{E}[M] &\le \frac{1}{t}\left(\log n + \frac{d}{2} \log (1-t)^{-1}\right),
    \end{aligned}
\end{equation}
where $MGF(t)_{M} = (1-t^2)^{-d/2}$ is the Moment Generating Function ({\sf MGF})~\cite{curtiss1942note} of distribution $M$, and $0<t<1$ is a tuning parameter for the tightness of estimation. According to Extreme Value Theory, $\mathbb{E}[M]$ can be estimated by $\frac{1}{t}\left(\log n + \frac{d}{2} \log (1-t)^{-1}\right)$. Given $\cos \theta=\frac{\langle x, y\rangle}{||x||||y||}$, $\theta$ can be estimated as $\arccos \left(\frac{1}{td}\left(\log n + \frac{d}{2} \log (1-t)^{-1}\right)\right)$. Given that $\langle x, y\rangle < d$, we can then approximate that $\cos \theta \approx \min \left (\frac{1}{td}\left(\log n + \frac{d}{2} \log (1-t)^{-1}\right), 1\right). $ When $n$ grows to $\infty$, $\cos \theta$ converges to 1 and $\theta$ converges to $0^\circ$.
\end{proof}

\noindent\textbf{Remark 3.2-Key takeaways from the above analysis:} 
\textbf{(1) Section \ref{sec:3-2-1}} shows Euclidean-based edge selection can effectively address the connectivity limitations of K-\textsf{NDG}, particularly for top-K retrieval. 
\textbf{(2) Section~\ref{sec:3-2-2}} demonstrates that Euclidean-oriented search inherently reduces angular distance, which aligns with one of the core objectives of \mips.

\begin{figure}[tb!]
\vspace{1ex}
\centering
\centerline{\includegraphics[width=\linewidth]{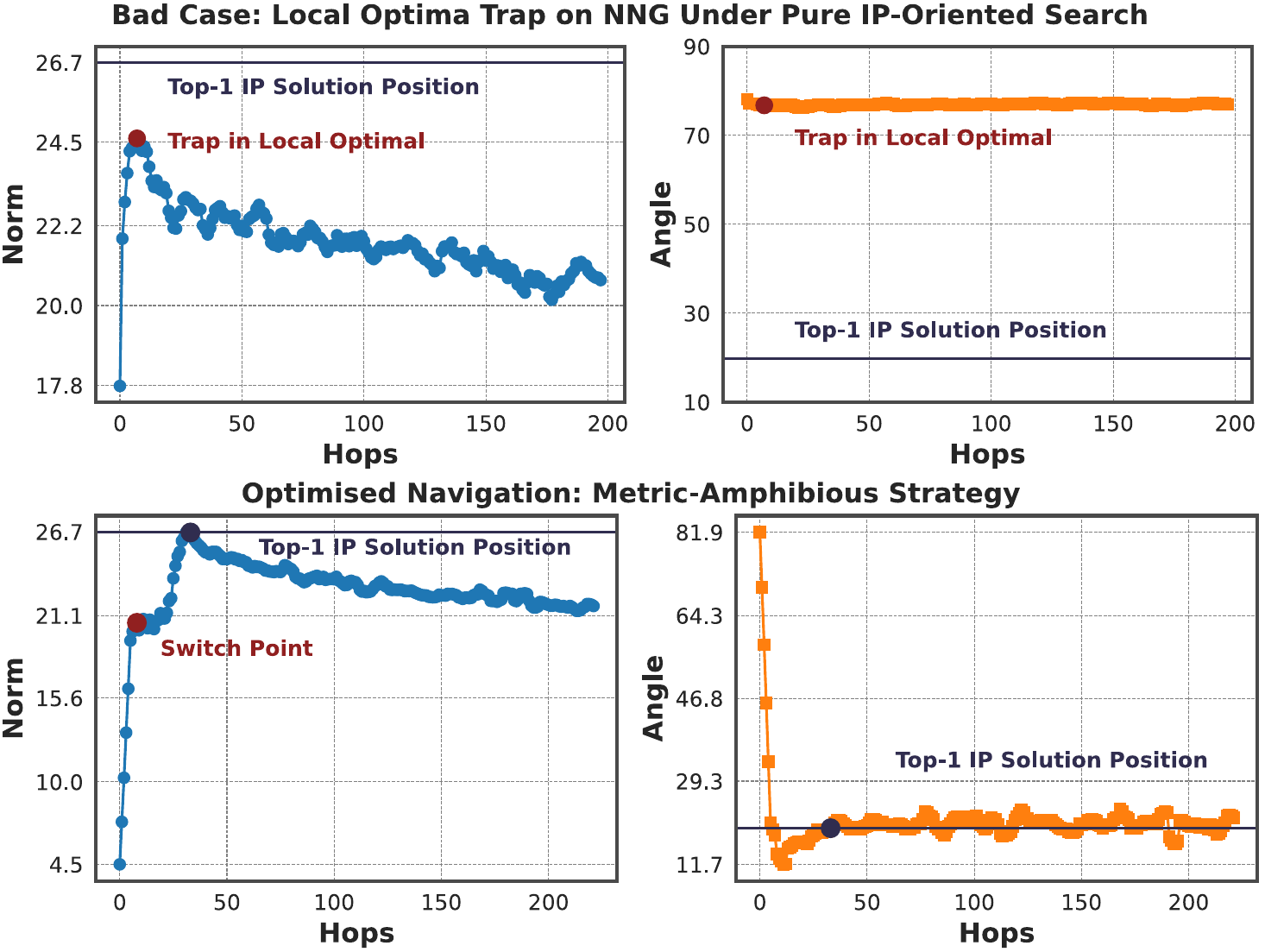}}
\vspace{-2ex}
\caption{\mips processes on Imagenet-1K Dataset through the lens of the average norm of the candidates, and the angles between the candidates and the query. The upper row illustrates a failure case when executing \mips directly on an \textsf{NSG}, where the algorithm fails to locate solutions even after 200 iterations. The lower row depicts the optimized search process using Metric-Amphibious strategies, which successfully identifies the top-1 solution within 50 iterations and achieves 100\% recall within 200 iterations.}
\vspace{-2ex}
\label{fig:bad_case}
\end{figure}

While Euclidean-oriented edges provide global connectivity and potential reachability to \mips solutions, a critical argument is that \textbf{ executing \mips solely on Euclidean-oriented graphs can be inefficient and prone to local optima traps}.

We conducted another preliminary test to evaluate \mips on \textsf{NNG}s (Figure~\ref{fig:bad_case}). Specifically, we run top-100 \mips on real-world datasets using \textsf{NSG}~\cite{fu2019fast}. 
Findings are: (1) \textsf{NNG}s achieve competitive recall (e.g., 98\% average recall on the \textsf{Imagenet-1k} dataset). (2) Certain queries receive $0$ recall, leading to a recall bottleneck.

Upon analyzing the failure cases on the \textsf{Imagenet-1K} dataset, we identified that these failures stem from suboptimal navigation: 
Without proper guidance, the search process maximizes the inner product inefficiently by alternating between norm expansion and angular minimization. This unordered navigation leads the search into local optima traps, as illustrated in the upper row of Figure~\ref{fig:bad_case}. 

Given the above analyses, we then tested the following \textbf{Metric-Amphibious strategies}: 
(1) Add $r$ edges from K-\textsf{NDG} for each node in \textsf{NSG}, linking them to the closet dominators; 
(2) Replace the first $m$ steps of \mips navigation with Euclidean-oriented navigation, prioritizing angular minimization within the candidate set $Q$ (Algorithm \ref{alg:gnns}). 
This Metric-Amphibious approach significantly improves recall, raising previously failed queries ($0$ recall) to 99\% recall within the same number of iterations. 
The lower row of Figure~\ref{fig:bad_case} demonstrates that, for the same query, the new strategy finds the top-1 solution within 50 iterations and locates 100\% of top-100 solutions within 200 iterations. In contrast, the upper bad case fails to find any solution even beyond 200 iterations.

%% file: 4-data.tex
\section{Methodology}
\label{sec:method}
The theoretical and empirical insights from Section~\ref{sec:analysis} highlight the potential of Metric-Amphibious Indexing and Search. Specifically: (1) Combining Euclidean- and \ip-based edge selection enhances graph connectivity and improves robustness in top-K retrieval; (2) Integrating Euclidean- and \ip-based search navigation reduces the risk of search trapping in local optima. Motivated by these findings, we propose a novel indexing and search framework below.

\subsection{Metric-Amphibious Graph}
\label{sec:indexing}
\begin{defn}[\textbf{\magg}]
\label{def:mag}
Given an MRNG~\cite{fu2019fast} $G$ constructed on a dataset $\mathcal{D}\subset R^d$, we extend it by identifying for each point $x\in\mathcal{D}$ $r$ dominator neighbors $q$ that maximizes $\langle x,q\rangle$, using the strategy in Definition~\ref{def:ndg}. The resulting graph $G'$ is referred to as the Metric-Amphibious Graph (MAG).
\end{defn}

\begin{theor}
\label{theorem:mag-properties}
An \magg $G$ defined on $\mathcal{D} \subset \mathbb{R}^d$ has the following properties: (1) $G$ is strongly connected; (2) the \mips answer $x^*$ of query $q$ is reachable from any starting point $x\neq x^*$ via a greedy search under the \ip metric (Algorithm~\ref{alg:gnns}); the amortized search complexity of \mips on $G$ is $O(\frac{cn^{1/d} \log (n)}{\Delta(n)})$, where $n=|\mathcal{D}|$, $c$ is a constant, and $\Delta(n)$ is a very slowly decreasing function of $n$ \citep{fu2019fast}.
\end{theor}
\begin{proof}
By Theorem 3 in~\cite{fu2019fast}, \textsf{MRNG} is strongly connected, ensuring at least one path between any node pair. The extended edges per node does not alter this connectivity, thus proving property (1).

By Theorem 1 in~\cite{fu2019fast}, \textsf{MRNG} guarantees a path between any two nodes can be found via Algorithm \ref{alg:gnns} under the Euclidean metric. Adding extra edges in \textsf{MAG} does not break this property. Moreover, by Fact~\ref{fact:scale_nn}, the \mips solution $x^*$ for query $q$ is reachable under the Euclidean metric by scaling $q$ to a hypothesis query $\mu q$. Due to this duality between \mips and \nns, this path can also be found guided by the \ip metric with Algorithm \ref{alg:gnns}, proving Property (2).

Following the same routine of proving Theorem 3 in \cite{fu2019fast}, we can prove \magg is an \textsf{MSNET}. By Theorem 2 in~\cite{fu2019fast}, the expected search path length in an \textsf{MSNET} follows: $\mathbb{E}[{L_{path}}]= \frac{n^{1/d} \log (n)}{\Delta(n)}$. Since \magg's out-degree is at most $R+r$ (where $R$ is the MRNG's maximal out-degree and can be treated as a constant given a fixed dimension $d$~\cite{fu2019fast}), the amortized search complexity of \mips on $G$ is $O(\frac{cn^{1/d} \log (n)}{\Delta(n)})$, where $c$ absorbs $d(R+r)$, proving Property (3).
\end{proof}
\vspace{-2mm}

\noindent \textbf{Key Takeaways}: By Theorem \ref{theorem:mag-properties}, \magg retains the connectivity strength of \textsf{MRNG} and benefits from the shortcuts to dominators from K-\textsf{NDG}, \textbf{enhancing navigation efficiency without affecting graph sparsity and without space transformation}.

Despite the theoretical advantages, a key limitation is the indexing complexity and memory efficiency. The standard construction of \magg incurs $O(DN^2)$ complexity, making it impractical for large datasets. Additionally, the high variance in out-degree among nodes may lead to memory inefficiencies. To address this, we propose a scalable approximation: (1) Construct \magg using approximate K-\textsf{NN} graphs (as an approximation for the Euclidean Delaunay graph); (2) Restrict the number of \textsf{MRNG}- and \textsf{NDG}-based edges, balancing efficiency and connectivity, detailed as follows.


\subsubsection{Two-Stage Construction Algorithm of \magg.}
By Fact~\ref{fact:scale_nn}, the K-\textsf{MIP} solutions for a query $q$ can be derived from a Euclidean proximity graph. However, obtaining $q$'s Euclidean neighbors from an \ip-oriented graph is inefficient due to potential low connectivity (Section~\ref{sec:domination}). Based on this, we propose the following pipeline.

\noindent\textbf{Stage 1: K-MRNG Approximation.} According to~\cite{fu2019fast}, an approximate \textsf{MRNG} can be constructed from a K-\textsf{NNG}. For each point $x\in\mathcal{D}$, we use \textsf{IVF-PQ}~\cite{faiss} to retrieve its $K$ nearest Euclidean neighbors, forming a K-\textsf{NNG}. Then we use \textsf{MRNG}'s edge selection strategy~\cite{fu2019fast} to sparsify this K-\textsf{NNG}. Finally, we retain only the closest $K_1$ neighbors, forming the approximate K-\textsf{MRNG}.

\noindent\textbf{Stage 2: K-MAG Approximation.} Utilizing Fact \ref{fact:scale_nn}, we execute Algorithm \ref{alg:gnns} using \ip metric to obtain K-\textsf{MIP} neighbors for each point $x\in\mathcal{D}$ on above K-\textsf{MRNG}. Then we apply \textsf{NDG}'s edge selection strategy (Definition \ref{def:ndg}) to sparsify the K-\textsf{MIP} neighbors. Finally, we retain only the closest $K_2$ \ip-oriented neighbors and inject them into K-\textsf{MRNG}, forming the approximate K-\textsf{MAG}.

\subsubsection{Metric-Amphibious Index Loading} 
To improve adaptability and memory efficiency, not all $K_1 + K_2$ edges are used during the search phase. 
Instead, we leverage the parameter $R$ to control the maximum out-degree of the Metric-Amphibious Graph (\magg) and the parameter $\alpha \in (0,1)$ to control the proportion of \ip-oriented edges. 
Specifically, $\alpha R$ \ip-oriented edges and $(1 - \alpha)R$ Euclidean-oriented edges are dynamically loaded, optimizing performance for different data distributions and varying $K$ in top-K retrieval.

\stitle{Indexing complexity.} The overall complexity consists of two parts: 
(1) Build a K-\textsf{MRNG} and (2) inject \textsf{NDG} edges. 
Let $n$ denote the number of vectors in $\mathcal{D}$, $R$ the maximum out-degree, and $d$ the dimension. 
The time complexity for the first step has an empirical complexity $O(ndR \log n)$~\cite{fu2019fast}. 
The second step involves two phases, where the search phrase has a complexity of $O(c_1n^{1/d}\log n)$ \cite{fu2019fast} and the pruning incurs $O(c_2d)$ per point. $c_1, c_2$ are constants. 
Overall, $O(ndR\log n)$ dominates the indexing time complexity.

\begin{figure}[tb!]
\centering
\centerline{\includegraphics[width=0.87\linewidth]{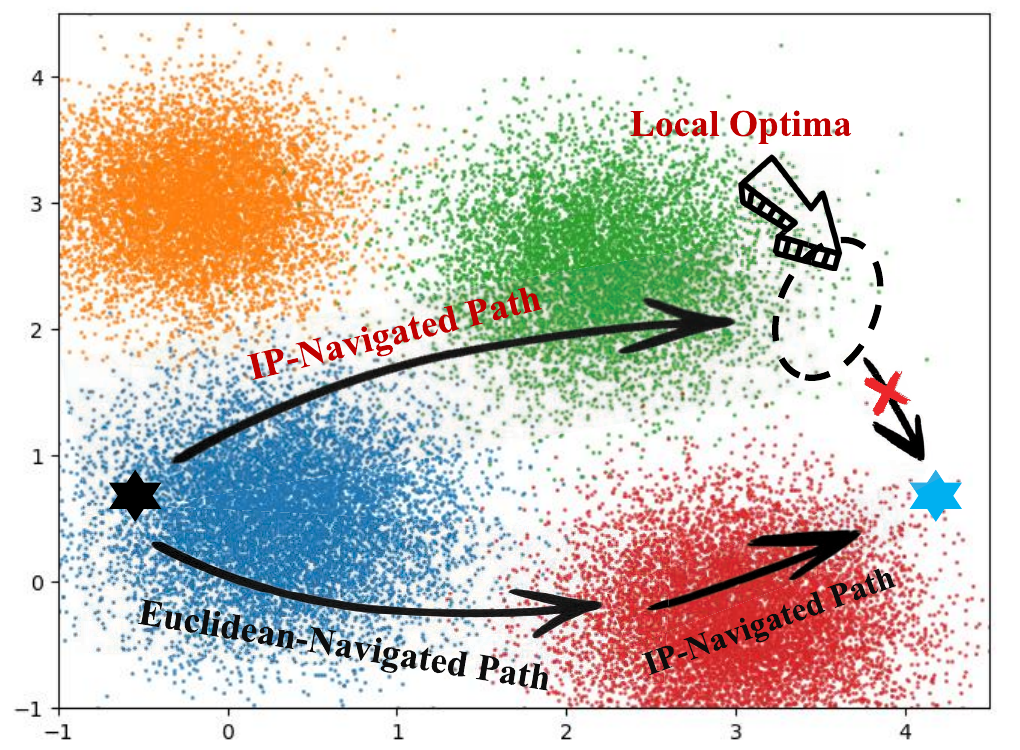}}
\vspace{-2ex}
\caption{Illustration of the impact of clustering. Without proper constraints, \mips tends to scale the norm or minimize the angle randomly during \ip maximization, depending on the data distribution and starting point. This process often traps the search in large-norm, sparsely populated regions (local optima). By incorporating Euclidean-oriented edges and improved navigation, this issue can be mitigated.}
\vspace{-2ex}
\label{fig:cluster-local-optima}
\end{figure}

\subsection{Adaptive Navigation with Metric Switch}
\label{sec:search}
Our analysis in Section~\ref{sec:analysis} highlights the benefits of combining different navigation strategies to prevent the search from getting trapped in local optima when maximizing \ip. 
This requires a minor modification to Algorithm~\ref{alg:gnns}. Formally, we propose \textbf{Adaptive Navigation with Metric Switch (ANMS)}, a two-stage search strategy that dynamically switches from Euclidean to \ip-based navigation. 
\textbf{The first stage} executes Algorithm \ref{alg:gnns} under Euclidean metric for $m$ steps. This minimizes the angular distances between the candidates and the query $q$; 
\textbf{The second stage} resumes Algorithm \ref{alg:gnns} upon the resulting candidates while switching the metric to maximize \ip. Here, $m$ is an extra tuning parameter to adapt \textsf{ANMS} to various data distributions, providing flexibility in controlling the initial search direction before switching metrics.

\textbf{The search complexity} of the ideal \magg is derived in Theorem~\ref{theorem:mag-properties}. While the approximate construction of \magg may slightly compromise this theoretical efficiency, our empirical evaluations in Section~\ref{sec:exp} (Figure~\ref{fig:complexity}) demonstrate that the search process still scales nearly as $O(\log n)$. 
These results confirm that the approximation introduced during construction has a minimal impact on the overall performance, ensuring that the search remains efficient.

\subsection{Distribution Aware Parameter Tuning}
In Section \ref{sec:indexing} and \ref{sec:search}, integrating Euclidean- and \ip-based strategies introduces two balancing parameters. 
This raises two important questions: 
\textbf{(1) How to tune the proportion ($\alpha$) of Euclidean- and \ip-based edges?} 
\textbf{(2) How to determine the metric switch position ($m$) in ANMS?} 
We find that these parameters are closely tied to the data distribution and retrieval quantity $K$, allowing us to derive key statistical indicators for efficient parameter tuning.

\subsubsection{High Sparsity of Dominators Favors \ip-Oriented Tuning}
When dominators are sparse, \textbf{IP-Oriented tuning} should be prioritized. This is because \mips solutions tend to concentrate in a smaller subset of the dataset, necessitating: 
\textbf{(1) Increase $\alpha$}, the fraction of \ip-oriented edges, to enhance direct connectivity to dominators; 
\textbf{(2) Reduce $m$}, the number of Euclidean-based search steps, to accelerate convergence to \mips solutions. Otherwise, the data favors Euclidean-oriented tuning, i.e., reducing $\alpha$ and increasing $m$.

As stated in Proposition~\ref{props:density}, the proportion of dominators in a dataset is highly dependent on the norm distribution of the vectors. 
Specifically, flatter norm distributions (with greater variability in norms) result in fewer dominators. 
In comparison, sharper norm distributions lead to a higher dominator proportion as it is harder for one point to dominate another with a similar norm. 
Another interpretation is that datasets with sharply distributed norms tend to approximate spherical distributions. In such cases, \mips approaches become closer to maximizing cosine similarity, which is inherently aligned with Euclidean-based \nns, thus favoring more Euclidean-oriented tuning.

To quantitatively guide parameter tuning, we introduce the Coefficient of Variation (\textsf{CV}) on the norm distribution, defined as $\sigma(\lVert x\rVert)/\mathbb{E}(\lVert x\rVert)$. Here $\sigma(\lVert x\rVert)$ is the standard deviation of vector norms. $\mathbb{E}(\lVert x\rVert)$ is the mean norm over the dataset. 

\noindent\textbf{Takeaways: High CV ($\ge 0.1$) indicates more \ip-oriented tuning, while low CV implies more Euclidean-oriented tuning.}

\subsubsection{Highly Clustered Data Favors Euclidean-Oriented Tuning}

The proposition~\ref{props:density} highlights the impact of dominator-oriented edges in highly clustered data, where their sparsity and concentration can lead to local optima traps, especially in inter-cluster regions. 
In such cases, inter-cluster connectivity depends on local Euclidean neighborhoods of points on cluster boundaries (Figure~\ref{fig:cluster-local-optima}). 
\textbf{Dominator-oriented edges are typically radial and may fail to bridge clusters}, causing suboptimal search paths. This issue can be mitigated using Euclidean-oriented edge selection strategies~\cite{fu2019fast,malkov2018efficient} and navigation~\cite{chen2023finger}. 
Thus, highly clustered datasets require more Euclidean-oriented tuning to enhance inter-cluster reachability.

To quantify the clustering, we employ the Davies-Bouldin Index (\textsf{DBI}), defined as: \textsf{DBI}$=\frac{1}{N}\sum_{i=1}^N\max_{j\neq i}\frac{\sigma_i+\sigma_j}{d(c_i,c_j)}$, where $N$ is the number of clusters, $\sigma_i$ is the average intra-cluster distance for cluster $i$, $c_i$ and $c_j$ are cluster centroids, and $d(c_i,c_j)$ is the distance between cluster centers. 
A higher \textsf{DBI} indicates lower inter-cluster separability. 
To capture clustering characteristics specific to \mips, we compute \textsf{DBI} under both Euclidean distance and cosine similarity, since Euclidean \textsf{DBI} captures density and spatial separation of clusters, while cosine \textsf{DBI} accounts for vector orientation variations, both of which are critical for \mips optimization.

\noindent\textbf{Takeaways: Highly clustered data (DBI ($\le 2$) under any distance metric) prioritizes Euclidean-oriented tuning, while evenly distributed data prioritizes \ip-oriented tuning.}

\subsubsection{Parameter tuning w.r.t. $K$} As shown in Section \ref{sec:domination}, smaller $K$ in top-K retrieval indicates the solutions are highly concentrated among dominators, where \ip-oriented tuning is preferred. Otherwise, Euclidean-oriented tuning is prioritized.

%% file: 6-exp.tex
\section{Experimental Evaluation}
\label{sec:exp}

In this section, we conduct extensive and comprehensive experiments to answer the following research questions: 
\textbf{RQ1:} How does \magg's search and indexing perform compared to existing methods? 
\textbf{RQ2:} How does \magg's metric-amphibious tuning contribute to its adaptability? 
\textbf{RQ3:} How does \magg behave at scale?

\begin{figure*}[tb!]
\centering
 \subfigure{\label{legend-qps-query}{
	\includegraphics[width=0.75\linewidth]{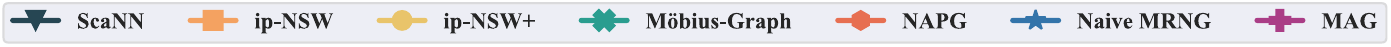}}}\\
\centerline{\includegraphics[width=0.95\linewidth]{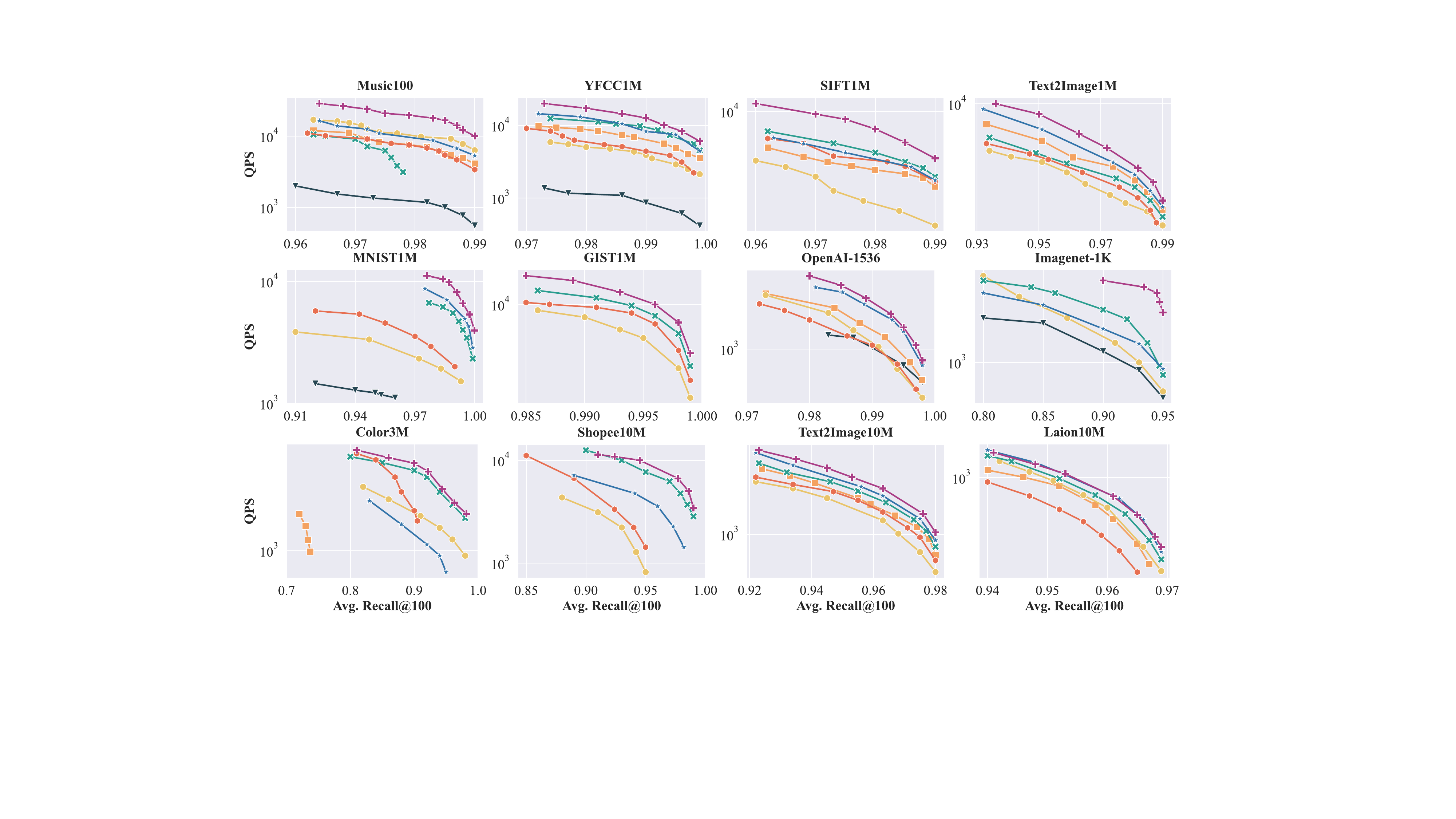}}
\vspace{-2ex}
\caption{Experimental results of search performance on real-world datasets. The upper right is better.}
\vspace{-0.9ex}
\label{fig:exp-qps}
\end{figure*}

\begin{table}[tb!]
  \caption{Dataset statistics, including the size of base and query data, dimensionality (Dim.), modality, DBI(Euclidean), DBI(Cosine), and CV~\cite{brown1998coefficient} of norm distribution.}
  \vspace{-3mm}
  \label{tab:prop}
  \resizebox{1\linewidth}{!}{
  \begin{tabular}{cccccccc}
    \toprule
    Dataset & Base & Dim. & Query& Modality & DBI(Euc.) & DBI(Cos.) & CV\\
    \midrule
    \textsf{Music100} & 1M & 100 & 10,000 & Audio &1.5&2.8&0.25 \\
    \textsf{YFCC1M} & 1M & 100 & 1,000 & Multi&1.51&2.9&0.07\\
    \textsf{SIFT1M} & 1M & 128 & 1,000 & Image&3.26&2.6&0.001\\
    \textsf{Text2Image1M} & 1M & 200 & 100,000  & Multi&2.5&3.0&0.03\\
    \textsf{MNIST1M} & 1M & 784 & 10,000 & Image&2.7&2.8&0.18\\
    \textsf{GIST1M} & 1M & 960 & 1,000 & Image&6.28&3.2&0.27\\
    \textsf{OpenAI-1536} & 1M & 1536 & 1,000 & Text&4.1&5.3&0.0\\
    \textsf{Imagenet-1k} & 1.3M & 1536 & 1,000 & Image &1&1.4&0.36\\
    \textsf{Color3M} & 3M & 282 & 1,000 & Image &2.6&2.1&0.17\\
    \textsf{Shopee10M} & 10M & 48 & 1,000 & E-commerce&2.4&2.1&0.24\\
    \textsf{Text2Image10M} & 10M & 200 & 100,000 & Multi&3.3 &3.6 &0.03\\
    \textsf{Laion10M} & 12M & 512 & 1,000 & Multi&4.3&3.6&0.0\\
    \bottomrule
  \end{tabular}}
  \vspace{-4mm}
\end{table}

\begin{table*}[tb!]  
  \caption{Experimental results on indexing time (s) and memory footprint (MB) of different methods on representative datasets.}
  \label{tab:index}
  \resizebox{\linewidth}{!}{\begin{tabular}
 {l *{8}{c} *{8}{c}}
    \toprule

    \cmidrule(lr){2-9} \cmidrule(lr){10-17}
 & \multicolumn{2}{c}{Music100} & \multicolumn{2}{c}{MNIST1M} & \multicolumn{2}{c}{GIST1M} & \multicolumn{2}{c}{OpenAI-1536} & \multicolumn{2}{c}{Imagenet-1k} & \multicolumn{2}{c}{Shopee10M} & \multicolumn{2}{c}{Text2Image10M} & \multicolumn{2}{c}{Laion10M}\\
\cmidrule(lr){2-3} \cmidrule(lr){4-5} \cmidrule(lr){6-7} \cmidrule(lr){8-9} \cmidrule(lr){10-11} \cmidrule(lr){12-13} \cmidrule(lr){14-15} \cmidrule(lr){16-17}
 & {Time} & {Memory} &{Time} & {Memory}   & {Time} & {Memory}   & {Time} & {Memory}  & {Time} & {Memory}  & {Time} & {Memory}  & {Time} & {Memory}  & {Time} & {Memory} \\
\midrule
\textsf{ScaNN} & 12 & 789 & 87 & 4666 & 106 & 5122 & 499 & 10240 & 1072 & 16820 & 67 & 7025 & 208 & 18480 & 1621 & 30656\\
\textsf{ip-NSW} & 68 & 584 & 106 & 3256 & 674 & 3921 & 721 & 6123 & 175 & 7580 & 431 & 4464 & 881 & 9656 & 4590 & 27146\\
\textsf{ip-NSW+} & 429 & 726 & 476 & 3588 & 1226 & 4233 & 1081 & 6206 & 990 & 7975 & 3217 & 5273 & 7722 & 11079 & 9728 & 28139 \\
\textsf{NAPG} & 84 & 648 & 526 & 3256 & 1086 & 3962 & 805 & 6123 & 820 & 7772 & 959 & 5452 & 1401 & 10250 & 6224 & 27146\\
\textsf{M{\"o}bius Graph} & 95 & 562 & 138 & 3256 & 930 & 3921 & 867 & 6123 & 338 & 7784 & 469 & 4464 & 1038 & 9720 & 3862 & 27146\\
\textsf{Naive MRNG} & 99 & 441 & 144 & 2960 & 792 & 3602 & 1334 & 5922 & 198 & 7505 & 667 & 2897 & 984 & 8426 & 2055 & 25882\\ 
\midrule
\textbf{\textsf{MAG (ours)}} & 134 & 496 & 230 & 2998 & 880 & 3660 & 1556 & 5990 & 219 & 7507 & 887 & 3194 & 1344 & 8795 & 2976 & 25920 \\
    \bottomrule
  \end{tabular}}
\end{table*}

\subsection{Experimental Setup}
\vspace{-1.5mm}

\stitle{Datasets.} We evaluate 12 real-world datasets with diverse cardinality, dimensionality, topology, and modality. Among them, \textbf{\textsf{Music100}}~\cite{morozov2018non}, \textbf{\textsf{MNIST1M}}~\cite{mnist}, \textbf{\textsf{Imagenet-1K}}~\cite{russakovsky2015imagenet}, \textbf{\textsf{SIFT1M}}~\cite{aumuller2020ann}, \textbf{\textsf{GIST1M}}~\cite{aumuller2020ann}, \textbf{\textsf{YFCC1M}}~\cite{thomee2016yfcc100m}, and \textbf{\textsf{Color3M}}~\cite{color3m} are widely used for the \mips problem. \textbf{\textsf{OpenAI-1536}}~\cite{dbpedia} is derived via the OpenAI text-embedding-3-large model on DBpedia, \textbf{\textsf{Shopee10M}}~\cite{fu2024residual} comes from a recommender system enhanced with advanced representation learning, \textbf{\textsf{Text2Image}}~\cite{text2image} and \textbf{\textsf{Laion10M}}~\cite{laion} provide cross-modality embeddings for cross-modal retrieval.

\stitle{Competitors.} We compare \magg against recent advanced methods of varying types: \textbf{(1) ip-NSW}~\cite{morozov2018non}, a graph-based approach utilizing an inner product navigable small world graph; \textbf{(2) ip-NSW+}~\cite{liu2020understanding}, which enhances \textsf{ip-NSW} by adding a high-tier angular proximity graph; \textbf{(3) M{\"o}bius Graph}~\cite{zhou2019mobius}, which utilizes the M{\"o}bius transformation to convert \mips to \nns; and \textbf{(4) NAPG}~\cite{tan2021norm}, an \ip native space method that introduces a norm-adjusted proximity graph. \textbf{(5) ScaNN}~\cite{guo2020accelerating}: a quantization method that integrates the recent state-of-the-art method \textbf{SOAR} \citep{sun2024soar} to further enhance performance. \textbf{(6) Naive MRNG}: the approximate K-\textsf{MRNG} constructed via Stage 1 solely as a strong baseline (Section \ref{sec:indexing}).

\stitle{Implementation.} All baselines are implemented in C++. The \textsf{ScaNN} library is called by Python bindings. Experiments are conducted on a single machine using 48 threads for index building across all methods. For query execution, we use same number of threads to ensure a fair comparison. Each experiment is repeated three times, and the average result is reported to reduce system variability. The code is at \textbf{\href{https://github.com/ZJU-DAILY/MAG}{https://github.com/ZJU-DAILY/MAG}}.

\stitle{Evaluation Protocol.} 
We evaluate the query performance using the common metric \textbf{Recall vs. Queries Per Second (QPS)}, which represents the number of queries an algorithm can process per second at each specified $recall@k$ level. The recall@k is defined as: $recall@k = \frac{|R \cap R'|}{|R|} = \frac{|R \cap R'|}{k}$, where $R$ is the ground-truth set of results, and $R'$ is the set of results returned by the algorithm. We use $k=100$, following prior practice. The memory footprint and indexing time are reported to evaluate indexing costs.

\subsection{Experimental Results}
\smallskip
\noindent\textbf{RQ1--Search}. Figure~\ref{fig:exp-qps} presents queries per second (\textsf{QPS}) against recall@100 for all methods. \textbf{Key findings: (1) MAG consistently outperforms all baselines} across datasets, due to its metric-amphibious framework and theoretical supports. Notably, It achieves 4$\times$ speedup over \textsf{M{\"o}bius Graph} on \textsf{Imagenet-1K}, 6$\times$ over \textsf{ip-NSW+} on \textsf{MNIST1M}, and 1.5$\times$ over \textsf{ip-NSW} on \textsf{Laion10M}, highlighting \magg's robustness. \textbf{(2) Baselines face accuracy bottlenecks} due to weak connectivity or local optima: \textsf{NAPG} struggles on 8 datasets, \textsf{ip-NSW} stagnates on 5 datasets, \textsf{Naive MRNG} caps at 90\% recall on \textsf{GIST1M}, and \textsf{M{\"o}bius Graph} fails at 80\% recall on \textsf{OpenAI-1536}. \textbf{(3) Data characteristics impact:} Low \textsf{CV} and high \textsf{DBI} datasets (\textsf{Laion10M} and \textsf{Text2Image10M}) favor Euclidean tuning. Thus \textsf{Naive MRNG} performs well besides \magg. High \textsf{CV} datasets (\textsf{Music100}, \textsf{MNIST1M}) favor \ip-tuning, where \magg stands out significantly. For low \textsf{DBI} datasets (\textsf{YFCC1M}, \textsf{Imagenet-1K}), balanced tuning gives \magg a clear advantage by efficient navigating across clusters. These align with our analysis. \textbf{(4) Dominator matters:} Despite strong connectivity, \textsf{Naive MRNG} underperforms as it fails to utilize concentration of \mips solutions among dominators.

\noindent\textbf{RQ1--Indexing}. Table~\ref{tab:index} summarizes the indexing time and memory usage for all competitors across eight representative datasets, omitting 4 due to space limit. \magg demonstrates moderate indexing and memory costs, similar on the 4 unreported ones. While \textsf{ScaNN} excels in fast indexing, it sacrifices search performance significantly. Overall, \magg achieves the best search-indexing trade-off.

\begin{figure}[tb!]
\vspace{1ex}
\centering
\centerline{\includegraphics[width=\linewidth]{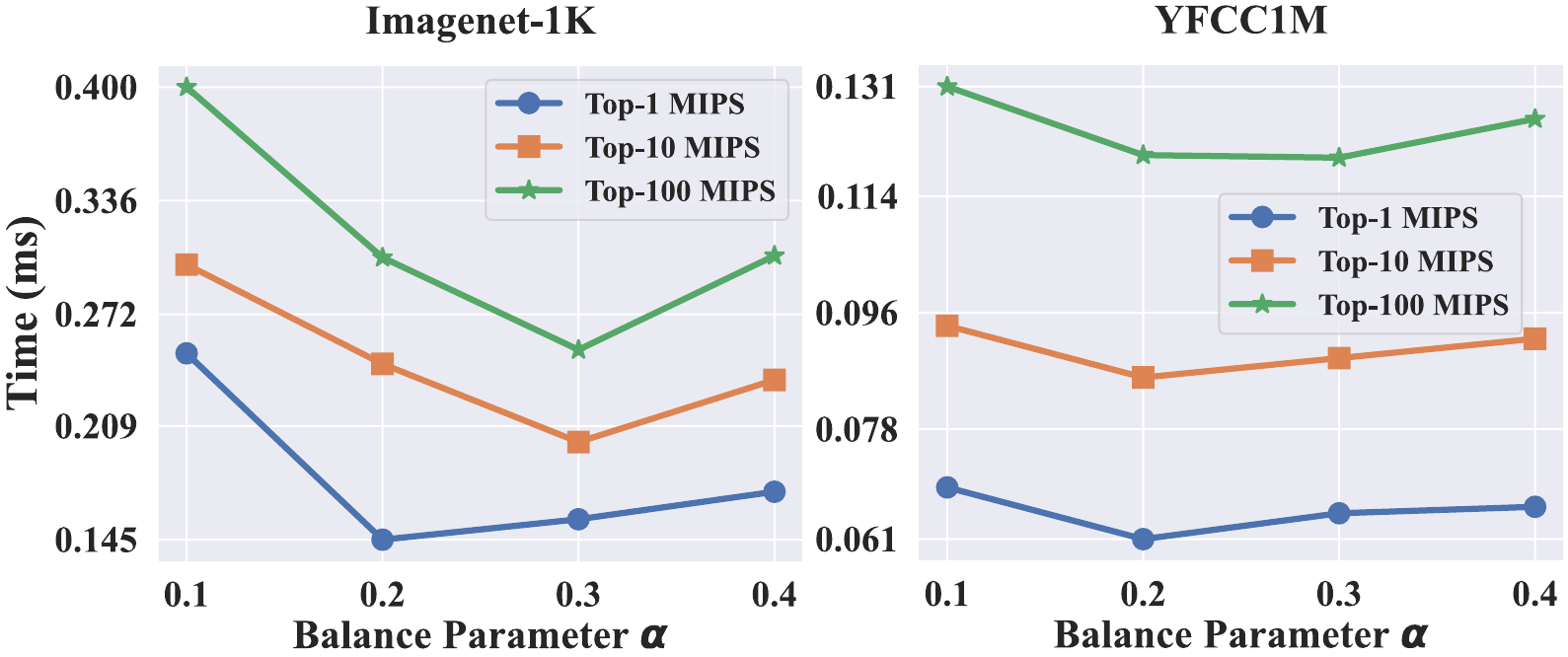}}
\vspace{-2ex}
\caption{Search time versus different \ip-edge ratio $\alpha$.}
\vspace{-4ex}
\label{fig:index-ratio}
\end{figure}

\noindent\textbf{RQ2: Metric-Amphibious Tuning.} We evaluate the impacts of Metric-Amphibious Tuning during the indexing and search phases over two representative datasets, regarding key parameters (1) the \ip-oriented edge ratio $\alpha$; (2) the metric switch position $m$.

\eetitle{Impact of $\alpha$.} We vary $\alpha$, which adjusts the ratio of \ip-oriented edges, and analyze its effect on search performance for top-K \mips ($K\in\{1,10,100\}$) on \textsf{Imagenet-1K} and \textsf{YFCC1M}. Results in Figure~\ref{fig:index-ratio} highlight: \textbf{(1)} Both \ip- and Euclidean oriented edges contributes to the efficiency of \magg, resulting in a concave curve for search performance. \textbf{(2)} Datasets with high \textsf{CV} favor larger $\alpha$, but excessive $\alpha$ reduces performance due to lower connectivity and local optima traps. Otherwise, low \textsf{CV} datasets favors smaller $\alpha$. \textbf{(3)} Excessively large $\alpha$ hurts the performance under larger $K$ more than smaller $K$.  

\eetitle{Impact of metric switch position $m$.} Fixing the indices, we vary $m$ and evaluate its effect on search speedup over the best competitor at 99\% recall. Table~\ref{tab:search-switch} shows: \textbf{(1)} Each dataset has a unique optimal $m$, driven by its specific data distribution. \textbf{(2)} Datasets with highly clustered data and sparse dominators benefit more from Euclidean-oriented navigation to avoid local optima traps. 

\begin{figure}[tb!]
\vspace{1ex}
\centering
\centerline{\includegraphics[width=\linewidth]{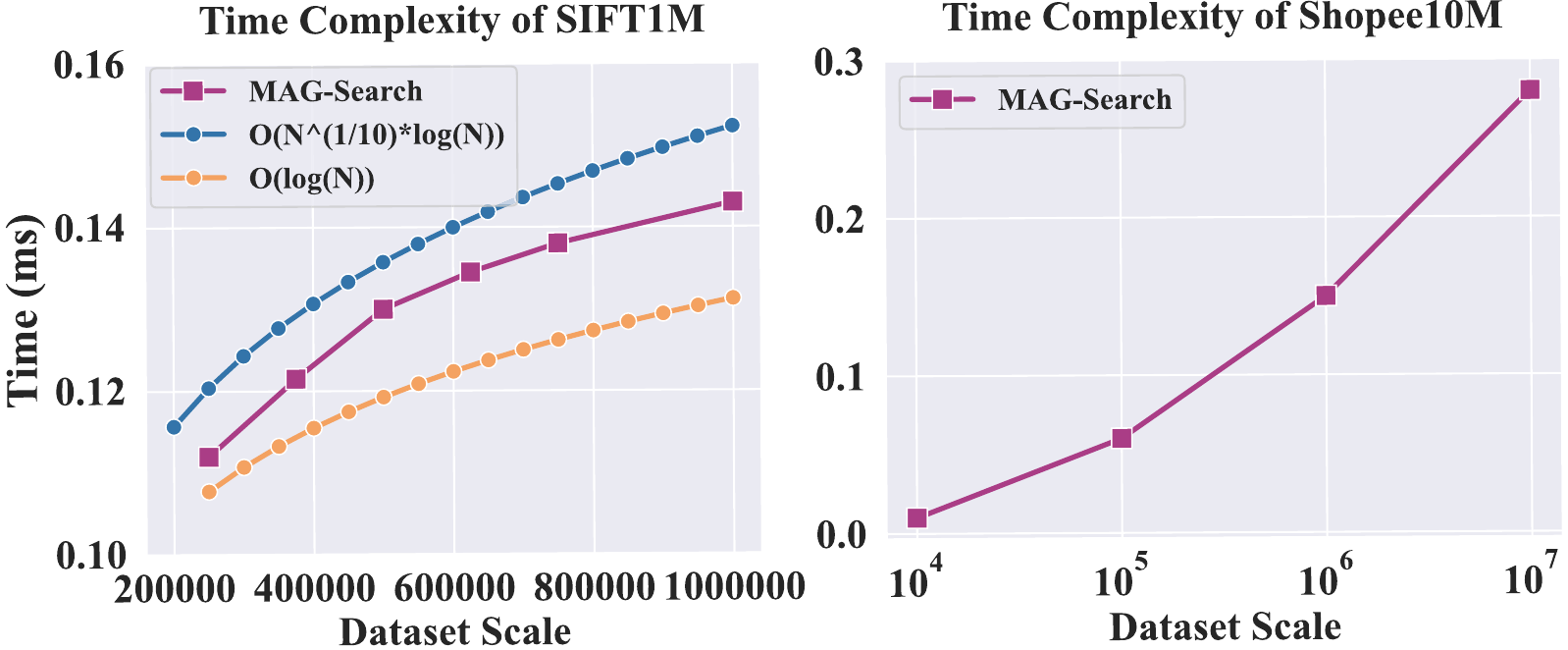}}
\vspace{-2ex}
\caption{Search time v.s.  scale on \textsf{SIFT1M} and \textsf{Shopee10M}.}
\vspace{-2ex}
\label{fig:complexity}
\end{figure}

\stitle{RQ3--Scalability.} We evaluate the search and indexing time complexity of \magg on \textsf{SIFT1M} and \textsf{Shopee10M}. Figure~\ref{fig:complexity} shows \magg’s search complexity scales near $O(\log n)$ on \textsf{Shopee10M} (more \ip-oriented), while slightly higher yet capped at $O(n^{1/d}\log n)$ on \textsf{SIFT1M} (more Euclidean-oriented). Note that the intrinsic dimension of \textsf{SIFT1M} is around 10. The indexing complexity of \magg follows near $O(n \log n)$ on both datasets, aligned with our analyses.

%% file: 7-related.tex
\section{Related Works}

\label{sec:related}
{Inner Product} is crucial in \textsf{AI} and machine learning applications such as representation learning,language modeling, computer vision and recommender systems~\cite{wang2024must,yu2014large,xu2020product,asai2023retrieval,huang2020embedding,radford2021learning}. \mips methods are generally categorized into Locality Sensitive Hashing (\lsh), tree-, quantization-, and graph-based approaches:

\stitle{LSH-based methods}: Traditional \lsh~\cite{wang2017survey,wei2024det}, originally designed for Euclidean space, is adapted for \mips using transformations such as $L_2$ \cite{shrivastava2014asymmetric}, Correlation \cite{shrivastava2015improved}, and \textsf{XBOX} \cite{bachrach2014speeding}. Range-\lsh~\cite{yan2018norm} is the first to observe that \mips results cluster around large-norm vectors. Simple-\lsh~\cite{neyshabur2015symmetric} introduce a symmetric \lsh that enjoys strong guarantees. Fargo \cite{zhao2023fargo} represents the recent state-of-the-art. 

\stitle{Tree-based methods}: Early \mips approaches favored trees but struggled with high dimensionality. \textsf{ProMIPS} \cite{song2021promips} addresses this by projecting vectors into a lower-dimensional space, though information loss remains a challenge. \textsf{LRUS-CoverTree}~\cite{ma2024reconsidering} improves on this but faces difficulties with negative inner product values. 

\stitle{Quantization-based methods}: \textsf{NEQ}~\cite{dai2020norm} quantizes the norms of items in a dataset explicitly to reduce errors in norm. ScaNN \cite{guo2020accelerating} integrates "VQ-PQ" with anisotropic quantization loss, while \textsf{SOAR} \cite{sun2024soar} employs an orthogonality-amplified residual loss and have become state-of-the-art and been integrated into ScaNN library. 

\stitle{Graph-based methods}: Proven effective for \nns, graph-based methods have been adapted for \mips. \textsf{ip-NSW}~\cite{morozov2018non} builds Delaunay graphs via inner product. \textsf{ip-NSW+}~\cite{liu2020understanding} improves graph quality with angular proximity. \textsf{M{\"o}bius-Graph}~\cite{zhou2019mobius} adopts M{\"o}bius transforms for \mips. \textsf{IPDG} prunes extreme points for top-1 MIPS. \textsf{NAPG}~\cite{tan2021norm} uses a norm-adaptive inner product ($\alpha \langle x, y \rangle$) in \textsf{ip-NSW}.

\begin{table}[tb!]
  \caption{The performance enhancement of \magg compared to the best competitor across various search switch steps on \textsf{Imagenet-1k} and \textsf{YFCC1M}.}
  \vspace{-2mm}
  \small
  \label{tab:search-switch}
  \resizebox{0.92\linewidth}{!}{
  \begin{tabular}{cccccc}
    \toprule
    Datasets & step=10 & step=20 & step=30 & step=40\\
    \midrule
    \textsf{Imagenet-1K} & 1.6x & 2.24x & 3.7x & 2.64x \\
    \textsf{YFCC1M} & 28\% & 39\% & 32\% & 25\%   \\
    \bottomrule
  \end{tabular}}
  \vspace{-2mm}
\end{table}

%% file: 8-conclu.tex
\section{Conclusion}
\label{sec:conclude}
This paper introduces a hybrid Metric-Amphibious framework for efficient and scalable \mips, including a novel graph index \magg and an efficient search algorithm \textsf{ANMS}. Comprehensive theoretical and empirical analysis support us to effectively leverage the strengths of both \ip- and Euclidean-oriented strategies while mitigating their limitations. Three statistical indicators sketching the data characteristics are identified to guide efficient parameter tuning. Extensive experiments demonstrate the efficiency, adaptability, and scalability of the proposed method.